\documentclass[showpacs,floatfix,twocolumn,byrevtex,longbibliography]{revtex4-1}

\usepackage[english]{babel}    % Hyphenation and whatnot
\usepackage{amsmath}           % Americal Math Soc. tools for math typing
\usepackage{graphicx}          % Including graphics in the document
\usepackage{float}             % Better handling of floats
\usepackage{bm}                % Bold symbol (math)
\usepackage[version=3]{mhchem} % Formula subscripts using \ce{}

\newcommand{\icm}{\ensuremath{~\textrm{cm}^{-1}}}% % cm-1 
\newcommand{\pol}[1]{\ensuremath{E \parallel #1}} % E // #1
\newcommand{\sig}[1]{\ensuremath{\sigma_1^{I_{#1}}}} % sigma A or C
\newcommand{\urusi}{\ce{URu2Si2}}
\newcommand{\tkl}{\ensuremath{T_{\it KL}}}
\newcommand{\tho}{\ensuremath{T_{\it HO}}}

\begin{document}

\title{Optical conductivity of \urusi\ in the Kondo Liquid and Hidden-Order Phases}
\author{R. P. S. M. Lobo}
\email{lobo@espci.fr}
\affiliation{ESPCI ParisTech, PSL Research University; CNRS; Sorbonne Universit\'es, UPMC Univ. Paris 6; LPEM, 10 rue Vauquelin, F-75231 Paris Cedex 5, France}

\author{J. Buhot} 
\altaffiliation[Present address: ]{High Field Magnet Laboratory, Institute for Molecules and Materials, Radboud University Nijmegen, Toernooiveld 7, 6525 ED Nijmegen, The Netherlands.}
\affiliation{Laboratoire Mat\'eriaux et Ph\'enom\`enes Quantiques, UMR 7162 CNRS, Universit\'e Paris Diderot, B\^at. Condorcet, 75205 Paris Cedex 13, France}

\author{M. A. M\'easson}
\affiliation{Laboratoire Mat\'eriaux et Ph\'enom\`enes Quantiques, UMR 7162 CNRS, Universit\'e Paris Diderot, B\^at. Condorcet, 75205 Paris Cedex 13, France}

\author{D. Aoki}
\affiliation{Universit\'e Grenoble Alpes, INAC-SPSMS, F-38000 Grenoble, France}
\affiliation{CEA, INAC-SX, F-38000 Grenoble, France}

\author{G. Lapertot}
\affiliation{Universit\'e Grenoble Alpes, INAC-SPSMS, F-38000 Grenoble, France}
\affiliation{CEA, INAC-SX, F-38000 Grenoble, France}

\author{P. Lejay}
\affiliation{CNRS, Institut N\'eel, F-38042 Grenoble, France}
\affiliation{Universit\'e Grenoble Alpes, Institut N\'eel, F-38042 Grenoble, France}

\author{C. C. Homes}
\affiliation{Condensed Matter Physics and Materials Science Department, Brookhaven National Laboratory, Upton, New York 11973, USA}

\date{\today}
\begin{abstract}
We measured the polarized optical conductivity of \urusi\ from room temperature down to 5~K, covering the Kondo state, the coherent Kondo liquid regime, and the hidden-order phase. The normal state is characterized by an anisotropic behavior between the \textit{ab} plane and \textit{c} axis responses. The \textit{ab} plane optical conductivity is strongly influenced by the formation of the coherent Kondo liquid: a sharp Drude peak develops and a hybridization gap at 12 meV leads to a spectral weight transfer to mid-infrared energies. The \textit{c} axis conductivity has a different behavior: the Drude peak already exists at 300 K and no particular anomaly or gap signature appears in the coherent Kondo liquid regime. When entering the hidden-order state, both polarizations see a dramatic decrease in the Drude spectral weight and scattering rate, compatible with a loss of about 50\% of the carriers at the Fermi level. At the same time a density-wave like gap appears along both polarizations at about 6.5 meV at 5 K. This gap closes respecting a mean field thermal evolution in the \textit{ab} plane. Along the \textit{c} axis it remains roughly constant and it ``fills up'' rather than closing. 
\end{abstract}
\maketitle
%

%%%%%%%%%%%%%%%%%%%%%%%%%%%%%%%%%%%%%%%%%%%%%%%%%%%%%%%%%%%%%%%%%%%%%%%%%%%%%%%
%
% Introduction
%
\section{Introduction}
\label{intro}
A consequence of the open $5f$ shells of Uranium is an array of electronic properties with the same energy scale. Competition and cooperation among these properties in intermetallic \urusi, in particular between itinerant and localized $f$ electrons, lead to a rich phase diagram  \cite{Mydosh2011,Mydosh2014,Schlabitz1986,Palstra1986,Maple1986}. \urusi\ is a heavy fermion material with a Kondo temperature of 370~K \cite{Schoenes1987}. At lower temperatures, hybridization between heavy $f$ electrons with conduction electrons creates a crossover to a Kondo liquid state \cite{Yang2008,Dordevic2001} having coherent transport properties below $\tkl \approx 70$~K. Upon further cooling, a second order mean-field transition at $\tho = 17.5$~K creates an electronically ordered state. The real nature of the order parameter remains unknown with the most varied hypothesis proposed \cite{Santini1994,Haule2009,Elgazzar2009,Pepin2011,Kusunose2011,Ressouche2012,Ikeda2012,Rau2012,Chandra2013,Thomas2013,Das2014,Riseborough2014}. Waiting for its elucidation the ``hidden-order'' moniker has been adopted to describe this phase \cite{Mydosh2011}. Closing the series of thermal phase transitions, an unconventional superconducting phase appears below $T_c \approx 1.5$~K \cite{Palstra1985}.  

The hidden-order transition is associated with a partial gap opening in both spin and charge channels \cite{Schoenes1987,Bonn1988,Fisher1990,Buhot2014,Buhot2015,Kung2015}. Recently, Shubnikov-de Hass \cite{Shishido2009,Hassinger2010,Altarawneh2011}, STM \cite{Schmidt2010,Aynajian2010}, and ARPES \cite{SantanderSyro2009,Yoshida2010,Kawasaki2011,Chatterjee2013,Boariu2013,Bareille2014} showed that \urusi\ has a complex Fermi surface with a strong renormalization below \tho. Point contact spectroscopy \cite{Hasselbach1992,Escudero1994,Thieme1995,Rodrigo1997}; ultra-fast pump-probe measurements \cite{Liu2011}; and optical conductivity \cite{Levallois2011} indicated the existence of a (pseudo)gap at temperatures ranging from 19~K to 30~K. In the case of point contact spectroscopy, it was suggested that the effect could be related to a local increase in \tho\ due to the pressure utilized to apply the contacts to the surface.

The unit cell of \urusi\ has a tetragonal symmetry with space group $I4/mmm$ ($D_{4h}^{17}$) \cite{Palstra1985} that seems to remain unchanged throughout the several phase transitions and transformations as a function of temperature, although this issue is still controversial \cite{Tonegawa2014,Tabata2014}. In a tetragonal structure, the optical properties are fully defined by two different tensor elements. Measurements with the electric field of light \pol{a} probe the \textit{ab} tetragonal basal plane. When \pol{c}, the measurements reveal the $c$ axis properties.

The early optical conductivity measurements of \urusi\ \cite{Bonn1988,Degiorgi1997} assumed an antiferromagnetic (AFM) transition below \tho. Even if the nature of the transition has since became unknown, their findings were not strictly tied to an AFM picture and fit nicely into the quest to understand the hidden-order character.  \citeauthor{Bonn1988} \cite{Bonn1988} were the first to measure the \pol{a} electrodynamics of \urusi\  and found a peak in the optical conductivity in the hidden-order phase, which had the properties of an optical gap that partially limited the states at the Fermi surface. Later, \citeauthor{Degiorgi1997} \cite{Degiorgi1997} published a comprehensive set of \pol{a} optical conductivity data on \urusi\ in the framework of several other heavy fermion compounds. They observed the development of a Drude peak below \tkl; a clear far-infrared signature below \tho; and estimated a mass enhancement of 68 at low temperatures, although this number is larger than what was previously proposed \cite{Maple1986,Schmidt2010,Nagel2012}.

The last few years have seen the revival of the interest in the optical conductivity of \urusi. \citeauthor{Levallois2011} \cite{Levallois2011} started this new wave and were also the first to measure the electrodynamics for \pol{c}. They essentially analyzed the response above \tho\ and found a suppression of spectral weight around 12~meV, below 30~K, in both $a$ and $c$ responses. They assigned this effect to a possible Fermi surface reconstruction crossover. \citeauthor{Guo2012} \cite{Guo2012} measured the \textit{ab} plane reflectivity and concluded that the pseudogap observed in Ref.~\cite{Levallois2011} was not a precursor of the hidden-order gap. \citeauthor{Hall2012} \cite{Hall2012} produced the first measurements along the $c$ axis in the hidden-order phase. Interestingly they found two gaps along the $c$ direction. One similar to its $a$ counterpart and another of smaller energy and magnitude. A gap at the same energy was also observed in the $A_{2g}$ symmetry by Raman spectroscopy \cite{Buhot2014,Kung2015}.

\citeauthor{Nagel2012} \cite{Nagel2012} showed that, just above \tho\ but still at low enough temperatures, 
the in-plane resistivity of \urusi\ had a quadratic dependence in both temperature ($T$) and frequency ($\omega$). They also showed that this property is not a sufficient condition to characterize a Fermi liquid as the ratio between the multiplying factors in $T$ and $\omega$ did not respect Landau's predictions.

In this paper we show the polarized optical conductivity of \urusi\ from room temperature though the Kondo liquid state and into the hidden-order phase. In Sec.~\ref{methods} we describe our samples and apparatus. Section \ref{results} describes qualitatively our measured reflectivity and optical conductivity. We discuss the Kondo liquid state in Sec.~\ref{kondo}, where we find a \pol{a} optical response strongly dependent on the Kondo physics with the formation of a coherent Drude peak and the opening of a hybridization gap at 12 meV below \tkl. Conversely, the \pol{c} data do not show any particular anomaly at \tkl\ and no signature of a hybridization gap. Section \ref{ho} analyses our results in the hidden-order phase. The strong band structure renormalization leads to a significant decrease of the Drude spectral weight as well as a large drop in the quasiparticle scattering rate. Both polarizations show the opening of a density-wave like gap at about 6.5 meV. The gap closes with a mean field behavior along \pol{a}. For the \pol{c} direction it fills up when approaching \tho\ with little temperature dependence. Our conclusions are summarized in Sec.~\ref{conclusions}.

%%%%%%%%%%%%%%%%%%%%%%%%%%%%%%%%%%%%%%%%%%%%%%%%%%%%%%%%%%%%%%%%%%%%%%%%%%%%%%%
%
% Methods
%
\section{Methods}
\label{methods}

Two single crystals were utilized in this study. The first crystal (sample A) was grown in a tri-arc furnace under argon atmosphere with a subsequent annealing under UHV at $900^\circ$C for 10 days. It had a surface containing the \textit{ab} plane of about $3 \times 6 \text{ mm}^2$ and it was about 200~$\mu$m thick. The second crystal (sample B) was grown by the Czochralski method using a tetra-arc furnace \cite{Aoki2010}. This sample was polished with a surface of $1 \times 1.5 \text{ mm}^2$ containing the \textit{ac} plane. It was further annealed under UHV at $950^\circ$C for two days in order to release the mechanical damage introduced by polishing \cite{Buhot2013}.

We utilized three different spectrometers to measured the near-normal incidence reflectivity from the far-infrared to the deep-UV: (i) a Bruker IFS113v from 2 meV (15\icm) to 12 meV (100\icm); (ii) a Bruker IFS66v from 6~meV (50\icm) to 2~eV ($15\,000\icm$); and (iii) an AvaSpec $2048 \times14$ optical fiber spectrometer from 1.5 eV ($12\,000\icm$) to 5~eV ($40\,000\icm$). We obtained the absolute value of the reflectivity with an \textit{in-situ} gold evaporation technique \cite{Homes1993b}. Our absolute accuracy is 0.5\% and the relative accuracy between different temperatures is better than 0.1\%. The spectral resolution was 2\icm (0.25 meV) in both Bruker spectrometers and 2 nm for the visible and UV ranges.

Measurements on sample A were taken on freshly cleaved surfaces and were restricted to the \pol{a} geometry. We did not use optical polarizers for this data set and we collected spectra at several temperatures from 5~K to 300~K from 2~meV (15\icm) to 2~eV ($15\,000\icm$). We extended this data to 5~eV ($40\,000\icm$) at room temperature only.

We utilized holographic wire grid polarizers on polyethylene and KRS-5 substrates to determine the \pol{c} response on sample B at several temperatures on a spectral range from 2~meV (15\icm) to 1~eV ($8\,000\icm$). We also measured the \pol{a} polarization for this sample at 5, 50 and 300~K in the far-infrared. We found the same results as the ones obtained in the cleaved sample A. 

Other optical functions were obtained from Kramers-Kronig analysis. At low energies we utilized a $1 - R \propto \sqrt{\omega}$ Hagen-Rubens extrapolation. For energies higher than our last temperature dependent point, we built a spectrum in four steps: (i) the \pol{a} 300~K data up to 5~eV ($40\,000\icm$); (ii) \citeauthor{Degiorgi1997} \cite{Degiorgi1997} data up to 12.5~eV ($100\,000\icm$); (iii) x-rays cross section reflectivity \cite{Tanner2015} up to 125~eV ($1\,000\,000$); and (iv) a free-electron $\omega^{-4}$ termination. We utilized this spectrum, at all temperatures, as the extension for the \pol{a} above 2~eV and, properly normalized, for \pol{c}. 

%%%%%%%%%%%%%%%%%%%%%%%%%%%%%%%%%%%%%%%%%%%%%%%%%%%%%%%%%%%%%%%%%%%%%%%%%%%%%%%
%
% Kondo state
%
\section{Results}
\label{results}

\begin{figure}[htb]
  \includegraphics[width=0.9\columnwidth]{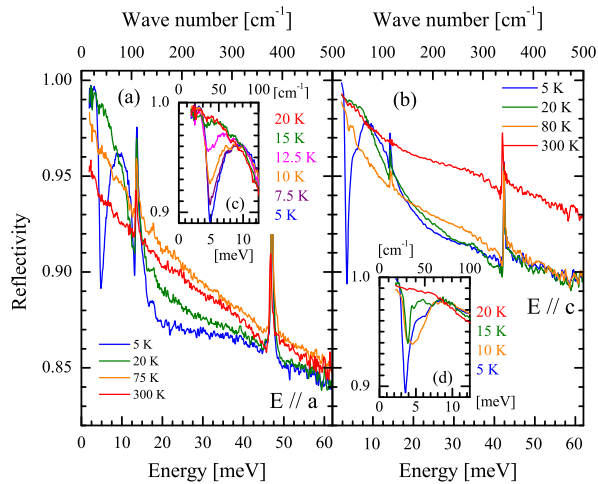} 
  \caption{(Color online) Reflectivity of \urusi\ above and below the hidden-order transition for (a) \pol{a} and (b) \pol{c} polarizations. Panels (c) and (d) detail the very far-infrared reflectivity in the hidden-order phase for both polarizations.}
  \label{fig1}
\end{figure}

Figure~\ref{fig1} shows the far-infrared reflectivity for both polarizations as a function of temperature. Panel (a) shows the \pol{a} response with a metallic profile at all temperatures. The sharp peaks at 15 and 45~meV are the two expected $E_u$ phonons \cite{Buhot2015}. The reflectivity increases steadily upon cooling the sample down to $\sim 75$~K. For lower temperatures, the reflectivity decreases below 45~meV and then increases again below 15~meV. In the hidden-order phase, a strong dip appears in the reflectance around 5~meV. The detailed temperature evolution of this hidden-order signature is depicted in panel (c). Note that its leading edge is constant but its trailing edge and amplitude are both strongly temperature dependent. 

Figure~\ref{fig1} (b) shows the \pol{c} reflectivity. When compared to the \pol{a} response, we note a few distinctions. Phonon frequencies are slightly different as they correspond to the two expected modes for the one-dimensional $A_{2u}$ representation \cite{Buhot2015}. The low-energy increase in the reflectivity is not as conspicuous. There is no clear temperature or energy marking a reflectivity decrease, we rather see a continuous evolution from 300 K. A similar hidden-order dip appears below \tho. However, as detailed  in panel (d), its temperature evolution is quite different.

\begin{figure}[htb]
  \includegraphics[width=0.9\columnwidth]{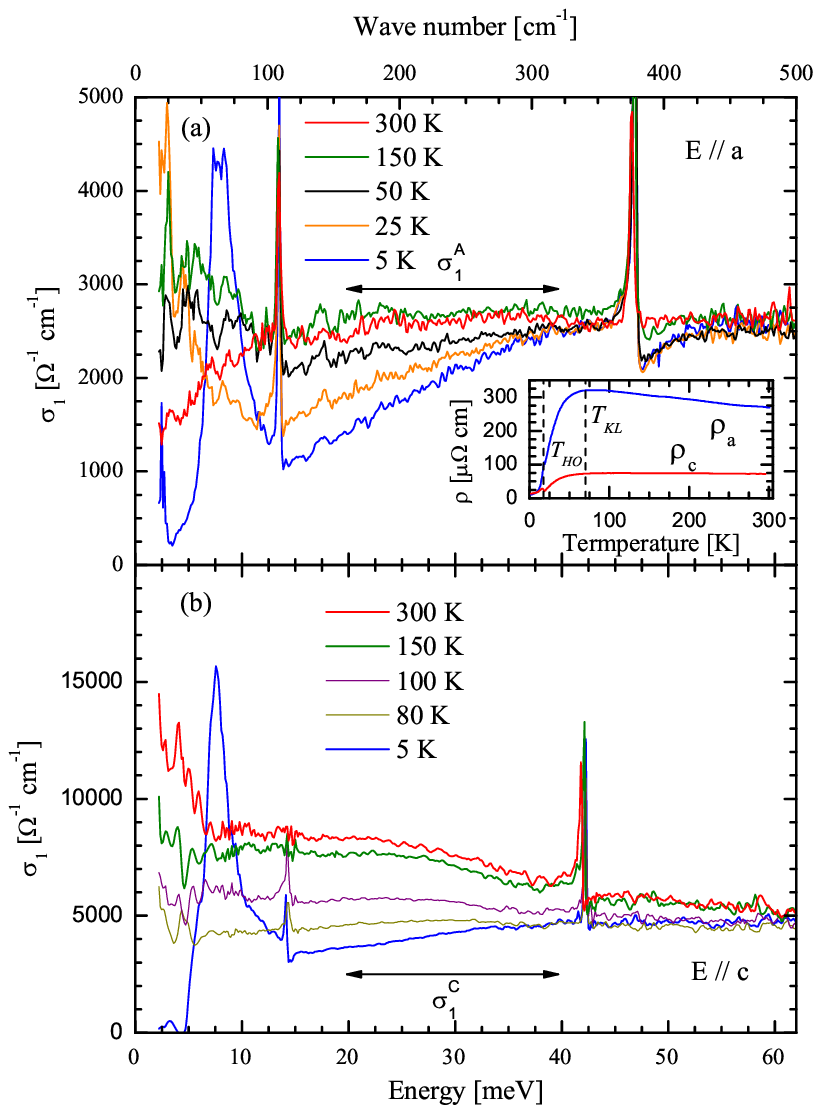} 
  \caption{(Color online) Optical conductivity of \urusi\ for (a) \pol{a} and (b) \pol{c}. The \pol{c} DC limit of $\sigma_1$ is roughly three times higher than its value for \pol{a} (note the difference in the optical conductivity scale), in accordance with the strongly anisotropic transport behavior of \urusi. In the 20--40~meV range we observe a decrease of spectral weight related to a narrowing of the Drude peak. For \pol{a} this effect happens mostly below the Kondo temperature. For \pol{c} the spectral weight decreases continuously from room temperature. Below the hidden order a strong peak appears at roughly 8~eV with a strong decrease of $\sigma_1$ at lower energies, characteristic of a partial gap at the Fermi level. The inset shows the resistivity from Refs.\cite{Palstra1986,Zhu2009,Nagel2012} (see text). The vertical dashed lines indicate the hidden-order transition and the boundary of the coherent Kondo liquid regime. }
  \label{fig2}
\end{figure}

Figure~\ref{fig2} shows the Kramers-Kronig obtained real part of the optical conductivity ($\sigma_1$). The sharp peaks around 15 and 45~meV are polar phonons and have been discussed extensively by \citeauthor{Buhot2015} \cite{Buhot2015}. Panel (a) shows $\sigma_1$ for \pol{a}. At 300~K, \urusi\ has an incoherent conductivity along $a$ without a clear Drude-like peak. In quite an opposite behavior, $\sigma_1$ actually shows a slight downturn at low frequencies. When cooling the material, one can discern a hint of a Drude peak at 150~K, which becomes well established upon further cooling down to 25~K. Below the hidden-order transition temperature, this Drude term collapses and a strong peak around 8~meV appears. This peak is directly related to the dip observed in the reflectivity.

Figure~\ref{fig2} (b) shows $\sigma_1$ for \pol{c}. Contrary to the \textit{ab} plane response, a Drude-like peak is already present at room temperature. The absolute value of the optical conductivity is roughly three times larger when compared to \pol{a} values, in accordance with the smaller dc resistivity. Upon cooling the sample, the optical conductivity below 40~meV decreases continuously, qualitatively suggesting a  Drude peak narrowing at low temperatures, as we will discuss in Sec.~\ref{kondo}. Once again, the hidden-order transition marks the appearance of a strong far-infrared peak with a significant redistribution of local spectral weight. 

%%%%%%%%%%%%%%%%%%%%%%%%%%%%%%%%%%%%%%%%%%%%%%%%%%%%%%%%%%%%%%%%%%%%%%%%%%%%%%%
%
% Kondo state
%
\section{Kondo Liquid State}
\label{kondo}

Let us first concentrate on the optical properties of \urusi\ at temperatures above the hidden-order transition. Besides the hidden-order signature, Fig.~\ref{fig2} shows large changes in the optical conductivity below $\sim 40$~meV. To discuss this effect and avoid complications due to phonons and the hidden-order feature, we will concentrate our analysis in the 20--40~meV region. We define \sig{a} and \sig{c} as the optical conductivity in this intermediate region for \pol{a} and \pol{c}, respectively.

\sig{a} remains almost temperature independent from 300~K down to about 75~K, roughly \tkl. Below this temperature, it decreases significantly whereas the low-energy conductivity increases with the development of a Drude peak. Conversely, along the $c$ axis the optical conductivity shows a far-infrared Drude-like peak at all temperatures and \sig{c} decreases steadily from 300~K. 

The inset of Fig.~\ref{fig2} (a) shows the resistivity of \urusi. Data for $\rho_a$ was measured for a specimen from the same batch as our sample A and is extracted from Ref. \cite{Nagel2012}. The low temperature ($T < 25$~K) $c$ axis resistivity is from Ref.~\cite{Zhu2009}. It was extended to higher temperatures with a proper renormalization of the data in Ref.~\cite{Palstra1986}.

We can qualitatively explain the optical conductivity behavior from the temperature dependence of the resistivity. As $\rho_c$ is smaller than $\rho_a$ it is reasonable to find a room temperature Drude peak in the former and not the latter. A temperature decrease leads to a slight increase in $\rho_a$ down to the coherent Kondo liquid temperature, explaining  why \sig{a} remains roughly constant down to \tkl. The dramatic drop in $\rho_a$  below \tkl\ is accompanied by the formation of a Drude peak in the \pol{a} polarization. The narrowing of this Drude peak leads to the observed decrease in \sig{a}. Along the \textit{c} axis, the relative decrease in $\rho_c$ is not as strong as the one observed along $a$. The steady decrease in \sig{c} can again be understood as a continuous Drude narrowing.

In heavy fermion compounds the Drude response is strongly renormalized by the effective mass \cite{Millis1987}, producing very narrow peaks \cite{Degiorgi1997,Scheffler2005,Dressel2006a}, difficult to access from the optical conductivity. For example, the low temperature scattering rate in \ce{UPd2Al3} is about 3~GHz (0.012~meV or 0.1\icm), a value more than two orders of magnitude smaller than our lowest measured frequency. The mass enhancement in \urusi\ is smaller but comparable to that in \ce{UPd2Al3} \cite{Dressel2006a} and one should also expect a narrow Drude peak. Indeed, $\rho_a^{-1} (T = 5 K) \approx 70\,000 \, \Omega^{-1} \icm$, an order of magnitude larger than the scale shown in Fig.~\ref{fig2} (a). This can only be explained by a very narrow Drude peak with a scattering rate much smaller than the lowest measured energy. Such narrow peaks are more easily seen in the frequency-dependent dielectric function, which can be parametrized by:
\begin{equation}
\varepsilon (\omega) = \varepsilon_{\it el} + \varepsilon_{\it ph} + \varepsilon_{\it HO} - \frac{\Omega_p^2}{\omega^2 + i \omega / \tau} \, ,
\label{eqDrude}
\end{equation}
where the first three terms on the right hand side are contributions from electronic transitions ($\varepsilon_{\it el}$); phonons ($\varepsilon_{\it ph}$); and the hidden-order ($\varepsilon_{\it HO})$. The last term is the Drude response from free carriers, characterized by a plasma frequency ($\Omega_p$) and a scattering rate ($1 / \tau$). We note that, when $1/\tau$ is small when compared to the lowest measured frequency, the real part of the dielectric function --- $\varepsilon_1(\omega)$ --- has a negative $\omega^{-2}$ divergence. The Drude model is a very crude single band approximation for mobile carriers in strongly correlated systems as it takes into account neither a frequency dependent scattering rate \cite{Allen1971,Puchkov1996a} nor multiband effects. However, it is a very good approximation for the low frequency response \cite{Romero1992}.

\begin{figure}[htb]
  \includegraphics[width=0.9\columnwidth]{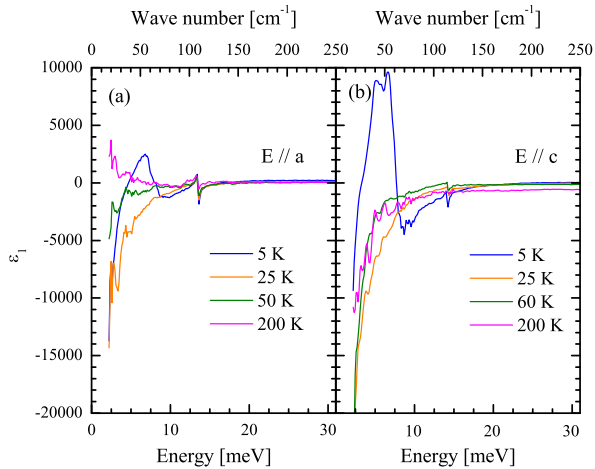} 
  \caption{(Color online) Real part of the dielectric function for (a) \pol{a} and (b) \pol{c}. At very low energies a negative divergence, due to the Drude contribution to Eq.~\ref{eqDrude}, dominates the spectral response and allows us to determine the plasma frequency and scattering rate. The peaks around 15 meV correspond to polar phonons. In the hidden-order state another large peak appears around 5 meV. Note that a well defined negative divergence is present at all temperatures for \pol{c} but only below 75~K for \pol{a}.}
  \label{fig3}
\end{figure}

Figure~\ref{fig3} shows $\varepsilon_1(\omega)$ for both polarizations. Panel (a) shows a low-energy positive upturn in $\varepsilon_1$ at 200 K for \pol{a}. This means that no coherent, well established Drude peak exists for this temperature. This situation persists down to temperatures close to \tkl. Only at 75~K (not shown) could we detect a slight hint of a negative low frequency $\varepsilon_1(\omega)$. This Drude signature gets stronger from \tkl\ down to \tho\ and remains present in the hidden-order phase. This means that, in agreement with the decreasing resistivity, the opening of a gap below \tho\ does not happen over the full Fermi surfaces, i.e., this is a partial gap. Panel (b) shows $\varepsilon_1(\omega)$ for \pol{c}. The remarkable difference with respect to the $a$ direction is that the Drude signature is present at all temperatures and not only below \tkl. 

\begin{figure}[htb]
  \includegraphics[width=0.9\columnwidth]{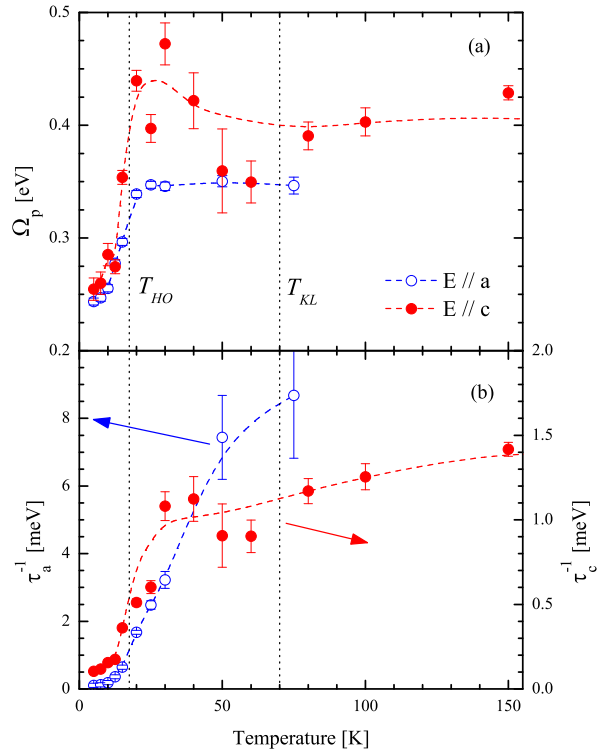} 
  \caption{(Color online) Drude parameters extracted from the low-energy real part of the dielectric function (Fig.~\ref{fig3}). Panel (a) shows the plasma frequency and panel (b) the scattering rate. There is no well defined Drude peak above $\sim 100$~K for data with \pol{a}. The vertical dotted lines indicates the position of \tho\ and \tkl. The dashed lines are guides to the eye. For clarity the data is only shown up to 150~k along \pol{c}. Above that temperature $\Omega_p$ is roughly constant and $1/\tau$ increases slightly. Error bars are estimated by utilizing different models for the hidden-order and phonon excitations as well as by varying the spectral range allowed for a least squares fitting of the Drude term.}
  \label{fig4}
\end{figure}

The presence of a Drude signature at low temperatures validates our picture of a very narrow Drude peak in $\sigma_1$ at low temperatures. The last term in Eq.~\ref{eqDrude} allows us to estimate the plasma frequency and the scattering rate of this narrow Drude component. The Drude parameters for both polarizations are shown in Fig.~\ref{fig4}. We will discuss the hidden-order phase in Sec.~\ref{ho}. Here we concentrate on the normal state.

As we mentioned before, in the \pol{a} polarization the Drude term only exists below \tkl. From that temperature down to \tho\ the plasma frequency is constant. No sign of a normal-state gap opening or band renormalization is present. The scattering rate decreases monotonically as for a regular metal. The \pol{c} polarization also shows a constant normal-state plasma frequency and, for this orientation, no effect of the Kondo transition is visible. Its scattering rate also decreases monotonically with temperature. 

As the low-energy Drude peak is not well defined at all temperatures (in particular for \pol{a}), we can attempt to infer the free carrier behavior from the optical conductivity in the intermediate regions \sig{a} and \sig{c}. For that, we define a restricted spectral weight, which measures the local charge distribution:
\begin{equation}
S(\omega_0,\omega_1) = \int_{\omega_0}^{\omega_1} \sigma_1(\omega) \, d\omega \, .
\label{eqrsw}
\end{equation}
When $\omega_0 \rightarrow 0$ and $\omega_1 \rightarrow \infty$, Eq.~\ref{eqrsw} yields $S = (\pi / 2) \varepsilon_0 \Omega_p^2$ leading to the standard $f$-sum rule. 

\begin{figure}[htb]
  \includegraphics[width=0.9\columnwidth]{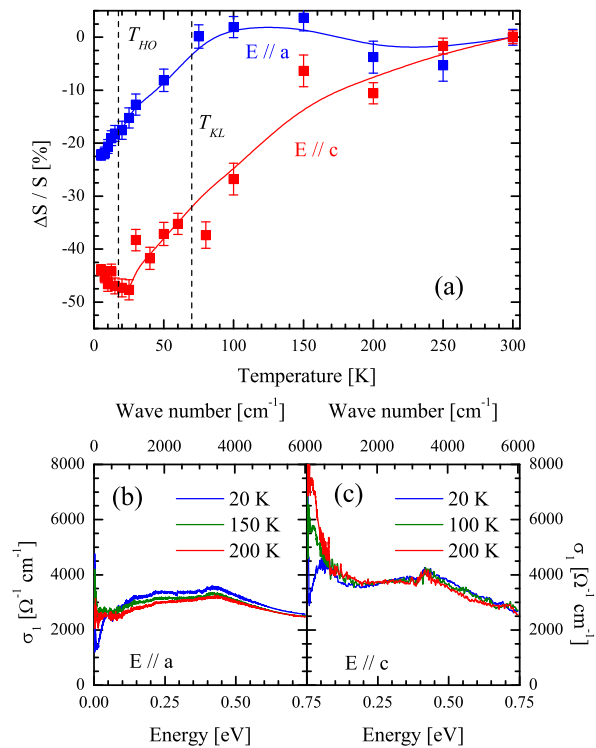} 
  \caption{(Color online) (a) Restricted spectral weight, normalized to its value at 300 K, calculated with $\omega_0 =  20$~meV (160\icm) and $\omega_1 = 40$~meV (320\icm). The solid lines are guides to the eye. The vertical dashed lines indicate the hidden-order transition and the boundary of the coherent Kondo liquid regime. Error bars are obtained by varying the cut-off frequencies by $\pm 10$\%. Optical conductivity for (b) \pol{a} and (c) \pol{c} in the mid-infrared range. Phonons were subtracted for clarity. For \pol{a}, part of the spectral weigh lost in the 20--40 meV range (see Fig.~\ref{fig2}) is transfered to the 0.5 eV region, indicating the opening of a gap above \tho. Along \pol{c} no such transfer is observed, within the accuracy of our data.}
  \label{fig5}
\end{figure}

We obtain Fig.~\ref{fig5} by setting $\omega_0 =  20$~meV (160\icm) and $\omega_1 = 40$~meV (320\icm), normalized by 300~K values. Along the $a$ direction we see that the total spectral weight in the intermediate region is roughly constant down to the Kondo liquid coherence temperature. This is a consequence of an almost incoherent transport as shown by the absence of a low-energy peak in the data. At \tkl\ the coherent far-infrared Drude peak appears. The integrated \sig{a} steadily decreases, indicating, as a first approximation, the narrowing of the Drude peak. This behavior corroborates the parameters found in Fig.~\ref{fig4} and indicate that the \pol{a} electrodynamics in the normal state is dominated by the Kondo physics.

The Drude parameters already hinted us that the \pol{c} polarization was not strongly influenced by the coherent Kondo liquid formation. Figure~\ref{fig5} confirms this observation by showing a continuous decrease of the total spectral weight in the intermediate region, with no features at \tkl. This decrease is compatible with the observed steady decrease, from room temperature, of $1 / \tau_c$ and a constant $\Omega_p$. 

It is important to note that the decrease in \sig{a} cannot be solely associated to the narrowing of the Drude peak. The decrease in the \sig{a} spectral weight below \tkl\ is not fully recovered in the Drude peak as extracted from $\varepsilon_1(\omega)$. Some of this spectral weight is transfered to the 0.5~eV region \cite{Guo2012}. This indicates that some of the spectral weight lost here is due to the hybridization gap opening. However, contrary to Refs.~\onlinecite{Levallois2011,Guo2012} we observe this gap opening in the vicinity of \tkl\ rather than 30~K. Note that Fig.~\ref{fig2} (a) shows this gap undoubtfully present at 50 K.

We can see further support for a gap-like structure related to the spectral weight decrease in \sig{a} by looking at its evolution in the hidden-order phase. The small scattering rate in the hidden-order phase, as shown in Fig.~\ref{fig4}, suggests that one should not expect any Drude spectral weight in the intermediate region used to calculate $\Delta S$ of Fig.~\ref{fig4}. The fact that this spectral weight continues to decrease in the hidden-order phase, means that it is being transfered not to the Drude peak but to higher energies. This also corroborates the view of independent hybridization and hidden-order gaps. Further evidence for a gap in the normal state for \pol{a} can be inferred from Fig.~\ref{fig5}(b). It shows an increase in the mid-infrared (around 0.5 eV) concomitant with the decrease observed in the 20--40 meV region (Fig.~\ref{fig2}), indicating that a transfer from low to high energies is also present, besides the narrowing of the Drude peak. The determination of the exact gap energy from the optical conductivity is model dependent. Nevertheless, gap values can be inferred by minima or by inflection points in $\sigma_1(\omega)$. Figure \ref{fig2}(a) shows that $\sigma_1$ at $T = 25$~K has a minimum around 12~meV. This is the same energy observed in optics \cite{Levallois2011,Guo2012} and by other techniques \cite{Maple1986,Boariu2013,Chatterjee2013}.

Conversely, along \pol{c}, we observe a saturation of $\Delta S$ in the hidden-order phase, when $1/\tau_c$ collapses. There is no sign of spectral weight transfer to higher frequencies, suggesting the absence of a hybridization gap in this direction. Indeed, inspection of Fig.~\ref{fig2}(b) does not reveal any clear minimum in $\sigma_1$ for \pol{c} in the normal state, which could be associated to a gap energy. Figure \ref{fig5}(c) further supports the absence of a hybridization gap signature along \pol{c}. Within the accuracy of our data, we observe no changes in the mid-infrared \pol{c} conductivity when going through the Kondo liquid coherence temperature. There is no mid-infrared spectral weight gain associated to the decrease in the 20--40 meV range.

We can summarize our findings for the optical conductivity of \urusi\ above \tho\ as 
(i) the Kondo coherence response is anisotropic; 
(ii) the \pol{a} electrodynamics is strongly dominated by Kondo physics with the formation of a very sharp Drude peak below \tkl; 
(iii) the narrowing of the Drude peak does not account for all the spectral weight lost in the 30~meV region and part of it is due to the hybridization gap that opens below \tkl\ at $\sim 12$~meV;
(iv) this hybridization gap is observed along the \pol{a} polarization but not along \pol{c}; 
(v) the decrease in $\sigma_1$ in the 20--40 meV for \pol{c} is fully attributable to a narrowing of a Drude peak, which starts at room temperature;
(vi) we observed no influence of the coherent Kondo liquid formation in the \pol{c} optical conductivity; 
(vii) in contrast to Ref. \onlinecite{Levallois2011}, we did not observe any effect related to a possible pseudogap opening at 30~K.

%%%%%%%%%%%%%%%%%%%%%%%%%%%%%%%%%%%%%%%%%%%%%%%%%%%%%%%%%%%%%%%%%%%%%%%%%%%%%%%
%
% Hidden order
%
\section{Hidden order}
\label{ho}

\begin{figure}[htb]
  \includegraphics[width=0.9\columnwidth]{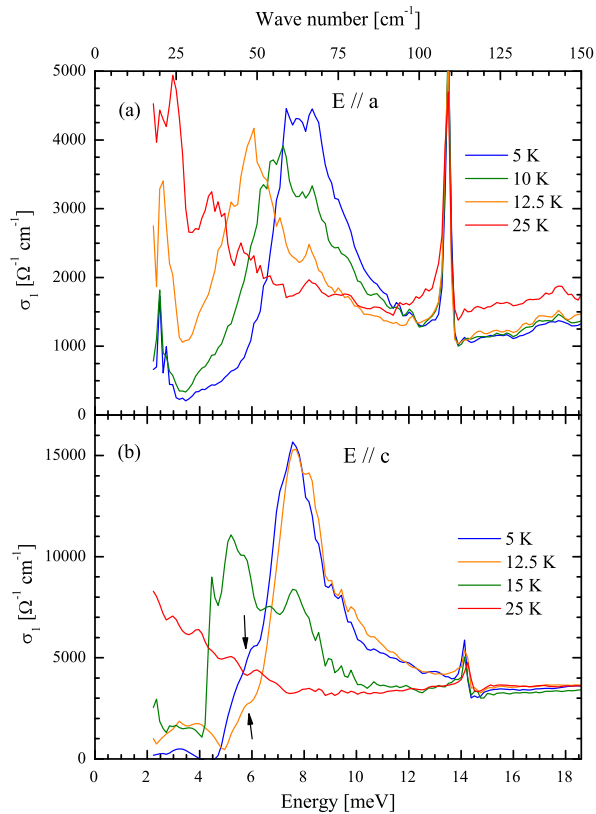} 
  \caption{(color online) Real part of the optical conductivity in the hidden-order phase for (a) \pol{a} and (b) \pol{c}. The large peak around 8~meV indicate the opening of a gap. In both cases, we can estimate (from the inflexion point) a gap $2\Delta_{HO} \approx 6.5$~meV at 5~K. The opening of the gap transfers spectral weight from low (just below $2\Delta_{HO}$) to high (just above $2\Delta_{HO}$) energies. Note that for \pol{a} the gap closes, \textit{i.e.}, its energy continuously decreases upon increasing temperature. For \pol{c} the gap fills up rather than close. Also note a small shoulder (indicated by the arrows) at the leading edge of the hidden-order peak along $c$, showing the presence of a second gap in this orientation. In both panels, the sharp peaks around 14~meV are polar phonons.}
  \label{fig6}
\end{figure}

When entering the hidden-order state, a sharp dip appears in the reflectivity, which leads to a peak in the far-infrared optical conductivity, as shown in Fig.~\ref{fig6}. The overall behavior for both polarizations is the same. A Drude-like peak dominates $\sigma_1$ just above \tho. In the hidden-order phase, a large peak appears around 8~meV at the expenses of a strong depletion in the low-energy $\sigma_1$. It is worth noting that in both cases, the spectral weight below 20~meV is conserved between the normal and the hidden-order state, \textit{i.e.}, the lost $\sigma_1$ at low energies corresponds to the area gained by the 8~meV peak. This ``low-to-high'' spectral weight transfer is a text-book example of a density wave gap (see chaps. 7 and 14 of Ref.~\onlinecite{Dressel2002}). 

One should take the ``density-wave'' terminology with caution. It simply means that the transfer of spectral weight is from below to above the gap energy, as opposed to a superconducting gap (transfer from finite energies to DC) or a semiconducting gap (gap developed over the whole Fermi surface with strictly zero absorption up to the gap energy). The gap behavior in \urusi\ does not mean that the hidden order is a density wave transition. However, it does impose that any model for the hidden-order transition must account for a gap with a spectral weight redistribution similar to that of a density-wave instability. 

The opening of a hidden-order gap comes from the strong band structure renormalization observed below \tho\ \cite{Chatterjee2013,Boariu2013}. However, in both polarizations, this gap is only partial. Indeed, the resistivity (inset of Fig.~\ref{fig2}) shows an increasing metallic character in the hidden-order state corresponding to the survival of a Drude peak in the dielectric function (Fig.~\ref{fig3}) down to our lowest measured temperature. Figure~\ref{fig4} (a) shows that the plasma frequency decreases dramatically at \tho, but does not vanish completely. The plasma frequency is related to microscopic quantities through $\Omega_p^2 = n e^2 / \varepsilon_0 m$. We can estimate the mobile charge density lost in the hidden-order state from the ratio $[\Omega_p^{(4K)} / \Omega_p^{(25K)}]^2$. We find that 50\% of the mobile carriers are lost in the \pol{a} direction and 60\% along $c$. This is in agreement with early results that estimated a partial gap opening over about 40\% of the Fermi surface \cite{Palstra1985,Maple1986}.

A closer analysis of the hidden-order gap shows remarkable differences between $a$ and $c$ responses. Figure \ref{fig6} (a) shows the \pol{a} spectra. Defining the exact gap energy is model dependent. A safe approach is to take the inflection point in the leading edge of the peak as a first (slightly overestimated) approximation of $2 \Delta_{HO}$, the energy necessary to excite a charge from the highest occupied state to the lowest empty state. From this perspective, panel (a) shows a hidden-order gap that decreases in both energy and spectral weight with the temperature, whereas panel (b) shows that the gap along the $c$ axis keeps the same spectral weight and energy up to, at least, $0.7 \, \tho$. In addition a small shoulder (indicated by the arrows) appears at the leading edge of the \pol{c} gap. This is the signature of a second gap along that direction as first observed by \citeauthor{Hall2012} \cite{Hall2012} and discussed in detail in that paper.

For the \pol{a} data, the spectral weight redistributed by the hidden order decreases continuously from 5 K, suggesting that states at the Fermi level are recovered fast when approaching $T_N$. In the \pol{c} polarization, the density of states lost at the Fermi level do not change until one gets close to \tho. This observation can also be inferred from Fig.~\ref{fig4} (a) where the value of the plasma frequency along $c$ drops, in the hidden-order phase, at a faster pace than in the \textit{ab} plane.

\begin{figure}[htb]
  \includegraphics[width=0.6\columnwidth]{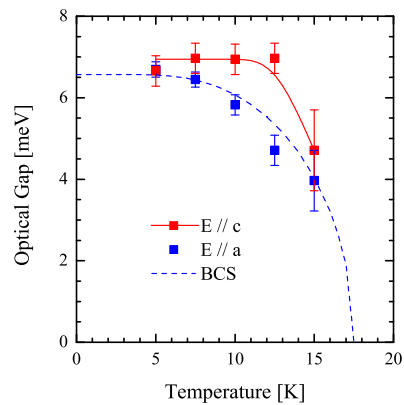} 
  \caption{(Color online) Temperature dependence of the hidden-order gap energy, taken as the inflexion point of the leading edge of the peak in Fig.~\ref{fig6}. Blue symbols correspond to the \pol{a} optical conductivity and red symbols to the \textit{c} axis. The dashed line is the result of a mean field BCS calculation and the solid line is a guide to the eye.}
  \label{fig7}
\end{figure}

Figure~\ref{fig7} shows the temperature evolution of the hidden-order gap for both orientations. Along the \pol{a} direction, the gap energy decreases with increasing temperature following a mean-field BCS-like behavior. Looking at Fig.~\ref{fig6} (a), one can also see that when the gap closes, the low frequency optical conductivity increases, indicating the presence of thermally excited quasiparticles. The $c$ axis behavior is qualitatively very different. The gap energy stays fixed up to temperatures very close to \tho. We do not find the quasi mean-field behavior observed in Ref.~\cite{Hall2012}. It does not close but rather ``fills-up''. In addition, the gap remains robust with almost no creation of low-energy thermally-excited quasiparticles almost all way up to \tho.

In the hidden-order phase a summary of our findings are: 
(i) passing the hidden-order transition leads to the opening of a density-wave like gap with spectral weight transfer from just below to just above the gap energy; 
(ii) the gap value is isotropic at 5~K but
(iii) its thermal evolution is very different between \pol{a} and \pol{c};
(iv) the temperature evolution of the gap energy along \pol{a} follows a BCS mean-field behavior but
(v) in the \pol{c} polarization it stays constant almost all the way up to \tho;
(vi) the \pol{a} gap closes when approaching \tho\ whilst the \pol{c} gap ``fills-up'' with almost no creation of thermally excited quasiparticles;
(vii) we confirm \citeauthor{Hall2012} \cite{Hall2012} observation of a second gap along the $c$ direction.

%%%%%%%%%%%%%%%%%%%%%%%%%%%%%%%%%%%%%%%%%%%%%%%%%%%%%%%%%%%%%%%%%%%%%%%%%%%%%%%
%
% Conclusions
%
\section{Conclusions}
\label{conclusions}

We measured the optical conductivity of \urusi\ in the normal and hidden-order state with \pol{a} and \pol{c} polarizations. Our results show strong anisotropic behaviors at all temperatures. In the normal state, we found an anisotropic behavior with respect to the Kondo liquid coherence formation. Along \pol{a} an incoherent optical conductivity is present down to \tkl\ where a sharp Drude peak develops. Conversely, along \pol{c} this Drude peak is already present at room temperature and shows no particular anomaly at \tkl. We observed the opening of a gap, due to the hybridization of localized $f$-electrons with mobile carriers, at temperatures very close to \tkl. It appears at an energy of 12 meV along the \pol{a} direction. We did not find a gap signature in the Kondo liquid state for \pol{c}. The Drude spectral weight of both polarizations decreases dramatically when entering the hidden-order phase and we estimate a loss of about 50\% of the carriers at the Fermi level. A fast drop in the scattering rate accompanies this decrease of spectral weight. A density-wave like gap appears along both polarizations in the hidden-order state at about 6.5 meV at 5 K. Along \pol{a} this gap closes following a mean field thermal evolution. Along \pol{c} the gap remains constant almost all the way up to \tho\ and it ``fills up'' rather than closing. 

\begin{acknowledgments}
%Thank you! Thank you very much!
We thank D.B. Tanner, T. Timusk, G. Kotliar, A. Millis, and C. P\'epin for fruitful discussions.
Work in Brookhaven National Laboratory is supported by the Office of Science, U.S. Department of Energy under Contract No. DE-SC0012704. The work in Universit\'e Paris Diderot was supported by the French Agence Nationale de la Recherche (ANR PRINCESS) and the Labex SEAM (Grant No. ANR-11-IDEX-0005-02).
\end{acknowledgments}

%%%%%%%%%%%%%%%%%%%%%%%%%%%%%%%%%%%%%%%%%%%%%%%%%%%%%%%%%%%%%%%%%%%%%%%%%%%%%%%
%
% The bibliography
%
\bibliography{biblio}

\end{document}